\documentclass[fleqn,usenatbib,useAMS]{mnras}

\usepackage[T1]{fontenc}

\DeclareRobustCommand{\VAN}[3]{#2}
\let\VANthebibliography\thebibliography
\def\thebibliography{\DeclareRobustCommand{\VAN}[3]{##3}\VANthebibliography}

\usepackage{graphicx}	% Including figure files
\usepackage{amsmath}	% Advanced maths commands
\usepackage{orcidlink}
\usepackage{comment}

\usepackage{newtxtext,newtxmath}
\title[Breakout from crystallization dynamos]{Magnetic field breakout from white dwarf crystallization dynamos}

\author[D. Blatman and S. Ginzburg]{
Daniel Blatman$^{\orcidlink{0009-0001-1957-8801}}$\thanks{E-mail: daniel.blatman1@mail.huji.ac.il} and
Sivan Ginzburg$^{\orcidlink{0000-0002-3751-4553}}$\thanks{E-mail: sivan.ginzburg@mail.huji.ac.il}
\\
Racah Institute of Physics, The Hebrew University, Jerusalem 91904, Israel
}

\date{Accepted XXX. Received YYY; in original form ZZZ}

\pubyear{2023}

\begin{document}
\label{firstpage}
\pagerange{\pageref{firstpage}--\pageref{lastpage}}
\maketitle

\begin{abstract}
A convective dynamo operating during the crystallization of white dwarfs is one of the promising channels to produce their observed strong magnetic fields. Although the magnitude of the fields generated by crystallization dynamos is uncertain, their timing may serve as an orthogonal test of this channel's contribution. The carbon--oxygen cores of $M\approx 0.5-1.0\,{\rm M}_{\sun}$ white dwarfs begin to crystallize at an age $t_{\rm cryst}\propto M^{-5/3}$, but the magnetic field is initially trapped in the convection zone -- deep inside the CO core. Only once a mass of $m_{\rm cryst}$ has crystallized, the convection zone approaches the white dwarf's helium layer, such that the magnetic diffusion time through the envelope shortens sufficiently for the field to break out to the surface, where it can be observed. This breakout time is longer than $t_{\rm cryst}$ by a few Gyr, scaling as $t_{\rm break}\propto t_{\rm cryst}f^{-1/2}$, where $f\equiv 1-m_{\rm cryst}/M$ depends on the white dwarf's initial C/O profile before crystallization. The first appearance of strong magnetic fields $B\gtrsim 1\textrm{ MG}$ in volume-limited samples approximately coincides with our numerically computed $t_{\rm break}(M)$ -- potentially signalling crystallization dynamos as a dominant magnetization channel. However, some observed magnetic white dwarfs are slightly younger, challenging this scenario. The dependence of the breakout process on the white dwarf's C/O profile implies that magnetism may probe the CO phase diagram, as well as uncertainties during the core helium burning phase in the white dwarf's progenitor, such as the $^{12}{\rm C}(\alpha,\gamma)^{16}{\rm O}$ nuclear reaction.
\end{abstract}


%%%%%%%%%%%%%%%%%%%%%%%%%%%%%%%%%%%%%%%%%%%%%%%%%%

%%%%%%%%%%%%%%%%% BODY OF PAPER %%%%%%%%%%%%%%%%%%

\section{Introduction}

White dwarf magnetic fields were first detected more than 50 years ago \citep{Kemp1970, AngelLandstreet71, LandstreetAngel71}, and strong fields of up to $\sim 10^9\textrm{ G}$ have been measured for many white dwarfs since \citep[see][for reviews]{Ferrario2015,Ferrario2020}. However, the origin of white dwarf magnetism remains an open question. One idea is that white dwarfs inherited their magnetic flux from previous stages of stellar evolution \citep{Angel1981,BraithwaiteSpruit2004,Tout2004,WickramasingheFerrario2005}. Alternatively, white dwarfs may have been magnetized by a dynamo operating during a common envelope event \citep{RegosTout1995,Tout2008,Nordhaus2011} or a double white dwarf merger \citep{GarciaBerro2012}. 

In a pioneering study, \citet{Isern2017} suggested that white dwarfs can drive magnetic dynamos as they evolve over time, even without interacting with a companion. When carbon--oxygen (CO) white dwarfs cool down sufficiently, their interiors begin to crystallize (solidify) from the inside out, with a crystallization front that gradually advances over several Gyr. The crystal phase is preferentially
richer in oxygen, which is drawn from the liquid CO mixture above the core, rendering the liquid unstable to compositional convection \citep{Stevenson1980,Mochkovitch1983,Isern1997}. Such convective flows may sustain a magnetic dynamo, in analogy to the Earth's iron--nickel core, where a similar compositional instability is thought to power the geodynamo \citep{Stevenson1983,MoffattLoper1994,ListerBuffett1995,GlatzmaierRoberts1997}. More massive oxygen--neon (ONe) white dwarfs may also harbour similar crystallization dynamos \citep{Camisassa2022}.

In a series of papers, \citet{Schreiber2021Nat,Schreiber2021,Schreiber2022,Schreiber2023} and \citet{Belloni2021} demonstrated that crystallization dynamos may potentially solve many of the outstanding puzzles of white dwarf magnetism, specifically, why some single and binary white dwarfs are magnetic while others are not. However, the velocity of the convective flows and the scaling laws that govern magnetic dynamos are currently both under debate \citep{Ginzburg2022,Fuentes2023,MontgomeryDunlap2023}. Therefore, the relative role of crystallization dynamos in explaining white dwarf magnetic fields is still unclear.

Here, we circumvent the uncertainties in the magnitude of the magnetic field, and focus instead on the orthogonal question of its timing. Crystallization begins at a specific age, when the white dwarf's core has cooled down sufficiently to transition from liquid to crystal in its centre. \citet{Bagnulo_2022} compared this age (as a function of the white dwarf's mass) to recent volume-limited surveys, in order to identify the magnetization channels of isolated white dwarfs. Their results indicate that crystallization dynamos may contribute significantly to the magnetism of old and cool white dwarfs, but there must be other channels as well, because some magnetic white dwarfs are too hot to be crystallizing \citep[see also][]{Ginzburg2022}. Specifically, \cite{Bagnulo_2022} argued that the most massive strongly magnetic white dwarfs probably resulted from double white dwarf mergers.   

However, the crystallization of the white dwarf's centre might not be an accurate indicator of the age when magnetic fields can be first observed. Crystallization dynamos generate magnetic fields only in a convective mantle that surrounds the crystal core. Magnetic fields have to diffuse from the edge of this convective zone to the white dwarf's surface to be detected, and the diffusion through the star's stable (i.e. non-convecting) outer layers can take several Gyr \citep{Isern2017,Ginzburg2022} -- comparable to the evolution time. Here, we compute this diffusion time as a function of the white dwarf's mass, and evaluate the age when magnetic fields break out to the surface, where they can be observed.

Our predicted breakout times provide a more precise observational test of the crystallization dynamo theory. Moreover, the magnetic diffusion time depends on the white dwarf's carbon and oxygen distribution. Therefore, combining our work with a large enough sample of magnetic white dwarfs may help to constrain the CO phase diagram during crystallization \citep[e.g.][]{BlouinDaligault2021}, as well as the $3\alpha$ and $^{12}{\rm C}(\alpha,\gamma)^{16}{\rm O}$ nuclear reaction rates during the core helium burning stage of stellar evolution, when carbon and oxygen are synthesized. 

The remainder of this paper is organized as follows. In Section \ref{sec:theory} we lay out the theoretical principles of the crystallization dynamo and the magnetic field's breakout. Section \ref{sec:scheme} describes our computational scheme. In Section \ref{sec:results} we present our computed breakout times, including the sensitivity to the phase diagram and nuclear reaction rates, and compare them to the observations. We summarize and discuss our findings in Section \ref{sec:summary}.

\section{Theory}\label{sec:theory}

\subsection{White dwarf cooling}

\citet{Mestel1952} derived a basic analytical theory of white dwarf cooling, which we briefly repeat here for completeness \citep[see also][]{VanHorn1971}. The pressure $P$ in the white dwarf's interior is dominated by degenerate electrons $P=K\rho^{5/3}$, where $K$ is a constant and $\rho$ is the density. In particular, the pressure at the centre (where the density is $\rho_{\rm c}\sim MR^{-3}$) balances the gravitational pressure
\begin{equation}\label{eq:prs_c}
    K\rho_{\rm c}^{5/3}\sim K\left(\frac{M}{R^3}\right)^{5/3}\sim\frac{GM^2}{R^4},
\end{equation}
where $M$ and $R$ are the white dwarf's mass and radius, and $G$ is the gravitational constant. We omit order-unity coefficients in this section with the goal of deriving approximate scaling relations. Equation \eqref{eq:prs_c} yields the white dwarf's mass--radius relation $R\propto M^{-1/3}$.

The efficient heat conduction of degenerate electrons implies that the white dwarf's interior is approximately isothermal, with a temperature $T_{\rm c}$. The pressure and density decrease towards the white dwarf's outer edge, until at some point the ideal gas pressure of the ions ($\propto\rho$) overcomes the electron degeneracy pressure ($\propto\rho^{5/3}$). We mark the density and pressure where the degeneracy is lifted by $\rho_0$ and $P_0$, which satisfy
\begin{equation}\label{eq:rho_0}
    P_0=K\rho_0^{5/3}\sim\frac{\rho_0}{\mu}k_{\rm B} T_{\rm c},
\end{equation}
where $\mu$ is the molecular weight and $k_{\rm B}$ is Boltzmann's constant. By combining equations \eqref{eq:prs_c} and \eqref{eq:rho_0}, we find that
\begin{equation}\label{eq:rho0_rhoc}
    \frac{\rho_0}{\rho_{\rm c}}\sim\left(\frac{k_{\rm B}T_{\rm c}R}{GM\mu}\right)^{3/2}\ll 1.
\end{equation}
The white dwarf's thin envelope from this point outwards can be treated as an ideal gas, and photon diffusion through this envelope regulates the cooling of the underlying isothermal core.
Using equations \eqref{eq:rho_0}, \eqref{eq:rho0_rhoc}, and the mass--radius relation, the optical depth of the envelope is given by
\begin{equation}\label{eq:tau}
    \tau\sim\frac{\kappa P_0}{g}\propto\frac{T_{\rm c}^{1/2}}{M^{5/3}},   
\end{equation}
where $g=GMR^{-2}$ is the surface gravity, and the opacity is given by Kramers' law $\kappa\propto \rho_0 T_{\rm c}^{-7/2}$. The white dwarf's luminosity is therefore
\begin{equation}
    L\sim\frac{R^2\sigma_{\rm SB} T_{\rm c}^4}{\tau}\propto MT_{\rm c}^{7/2},
\end{equation}
where $\sigma_{\rm SB}$ is the Stefan--Boltzmann constant.
The cooling time to a given central temperature is independent of the mass
\begin{equation}\label{eq:cooling_time}
    t\sim\frac{Mk_{\rm B}T_{\rm c}}{\mu L}\propto T_{\rm c}^{-5/2},
\end{equation}
with the white dwarf's heat capacity dominated by the non-degenerate ions.

\subsection{Crystallization}

When the white dwarf's interior cools down sufficiently, it begins to freeze into a crystal lattice. This phase transition occurs when the plasma coupling parameter (the ratio of the Coulomb to thermal energies)
\begin{equation}\label{eq:gamma}
    \Gamma=\frac{(ze)^2}{ak_{\rm B}T}\propto\frac{\rho^{1/3}}{T}
\end{equation}
exceeds a critical value $\Gamma_{\rm crit}\approx 200$ \citep{VanHorn1968,PotekhinChabrier2000,Bauer2020,Jermyn2021}; $z$ is the atomic number, $e$ is the electron's charge, $T$ is the temperature, and $a\propto\rho^{-1/3}$ is the average separation between ions. See \citet{Bauer2020} and \citet{Jermyn2021} for how to average $\Gamma$ for multiple ion species $z$. 

The white dwarf's centre has a density $\rho_{\rm c}\sim MR^{-3}\propto M^2$, and according to equations \eqref{eq:cooling_time} and \eqref{eq:gamma}, it crystallizes at an age  
\begin{equation}\label{eq:t_cryst}
    t_{\rm cryst}\propto T_{\rm c}^{-5/2}\propto M^{-5/3}.
\end{equation}
More external layers of the white dwarf's degenerate (and isothermal $T=T_{\rm c}$) core have a lower density $\rho<\rho_{\rm c}$, and therefore reach $\Gamma_{\rm crit}$ and crystallize at a later time. Quantitatively, we can generalize equation \eqref{eq:t_cryst} and calculate the time it takes the crystal core to reach a mass of $m_{\rm cryst}$. We denote by $\Delta m\equiv M-m_{\rm cryst}$ the mass that has not crystallized yet. From hydrostatic equilibrium, in the limit $\Delta m\ll M$ (which we adopt to derive a simple scaling, even though it is valid only at late times) the pressure at the crystallization front is given by
\begin{equation}
    P=K\rho^{5/3}\approx\frac{\Delta m g}{4\upi R^2}=\frac{GM\Delta m}{4\upi R^4}.
\end{equation}
The density there is given by comparing with equation \eqref{eq:prs_c}
\begin{equation}\label{eq:rho_f}
    \frac{\rho}{\rho_{\rm c}}\sim \left(\frac{\Delta m}{M}\right)^{3/5}=f^{3/5},
\end{equation}
where we assume that $\rho_0<\rho<\rho_{\rm c}$ (i.e. the crystallization front is inside the degenerate core) and define
\begin{equation}
    f\equiv\frac{\Delta m}{M}=1-\frac{m_{\rm cryst}}{M}.
\end{equation}
According to equation \eqref{eq:gamma}, the core has to cool down to a lower temperature $T_{\rm c}\propto\rho^{1/3}$ for this shell to crystallize, which is reached at an age $t>t_{\rm cryst}$ given by
\begin{equation}\label{eq:t_f}
    t=t_{\rm cryst}\left(\frac{\rho}{\rho_{\rm c}}\right)^{-5/6}=t_{\rm cryst}f^{-1/2}\propto M^{-5/3}f^{-1/2},
\end{equation}
where we also used equations \eqref{eq:cooling_time} and \eqref{eq:rho_f}. In this way, the crystal core gradually grows in mass $m_{\rm cryst}$ (i.e. $f$ decreases) as the white dwarf continues to cool. 

\subsection{Magnetic fields}

\begin{figure*}
	\includegraphics[width=\textwidth]{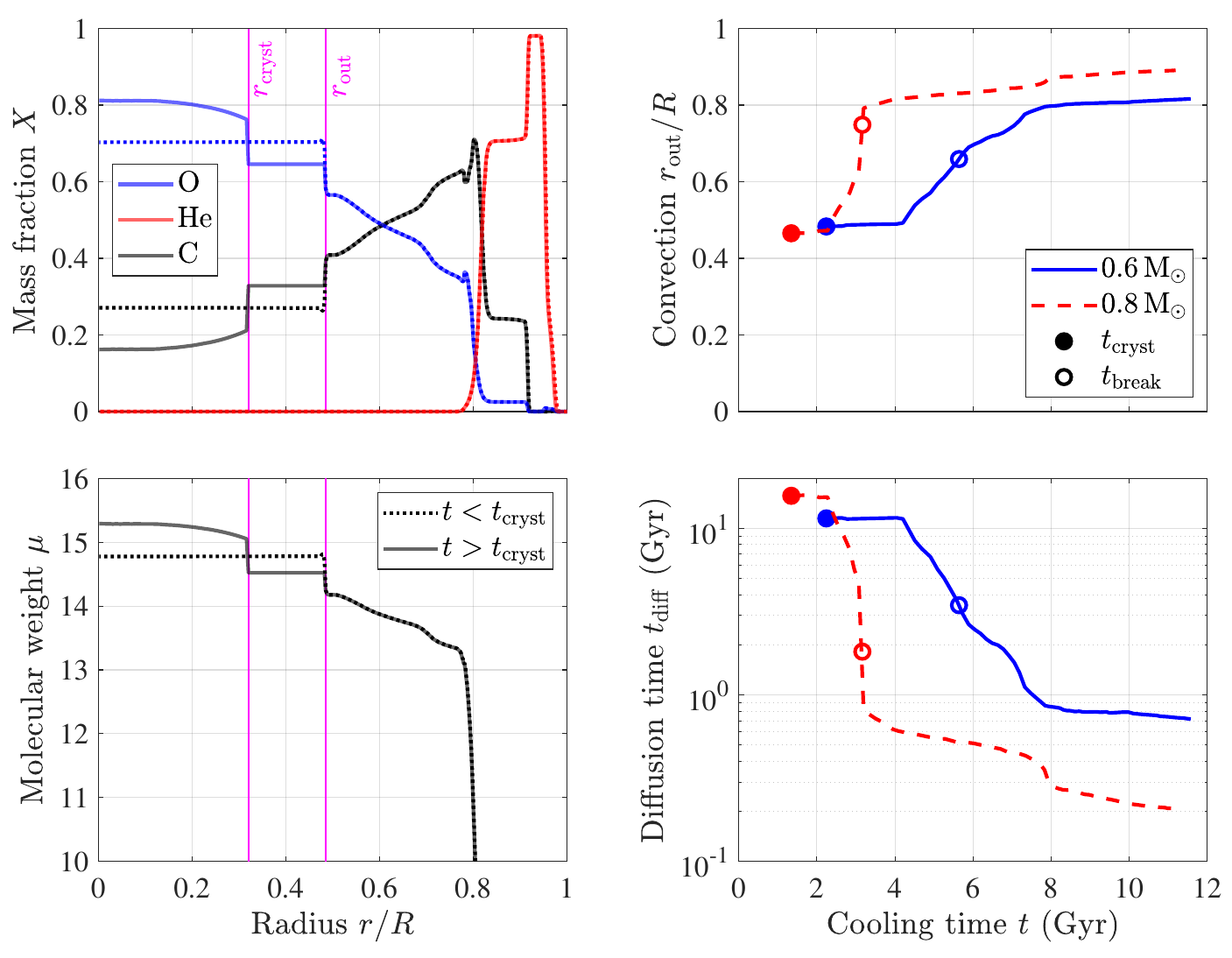}
    \caption{Breakout of the crystallization dynamo's magnetic field to the surface. \textit{Top left:} helium, carbon, and oxygen abundances inside a $0.6\,{\rm M}_{\sun}$ white dwarf before the onset of crystallization at $t_{\rm cryst}$ (dotted lines), and later (solid lines), when the crystal core has reached a radius $r_{\rm cryst}$. \textit{Bottom left:} the corresponding mean molecular weight $\mu$. Crystallization lowers $\mu$ of the liquid above the crystal core, rendering it unstable to convection up to an outer radius $r_{\rm out}$, where the ambient unperturbed $\mu$ is sufficiently low. Convection mixes the region $r_{\rm cryst}<r<r_{\rm out}$, and also sustains a magnetic dynamo there. \textit{Top right:} as crystallization progresses over time, a lower $\mu$ is required in order to stabilize the liquid, such that $r_{\rm out}$ expands. Most of the time, convection stops at a steep $\mu$ gradient. At early times $r_{\rm out}\approx 0.5 R$, corresponding to the jump in the C/O ratio of the original profile. Only at a later time convection reaches the white dwarf's atmosphere. \textit{Bottom right:} the magnetic diffusion time $t_{\rm diff}$ from $r_{\rm out}$ to the white dwarf's surface. The magnetic field breaks out to the surface at $t_{\rm break}=t_{\rm cryst}+t_{\rm diff}(t_{\rm break})$.}    \label{fig:theory}
\end{figure*}

An important feature of CO crystallization is a phase separation into an oxygen--enriched solid core and a carbon--enriched liquid mantle. This liquid layer above the growing core is lighter than the surrounding CO mixture, such that it rises up due to buoyancy, driving convection \citep{Stevenson1980,Mochkovitch1983,Isern1997}. We demonstrate the phase separation and convection in Fig. \ref{fig:theory}: crystallization lowers the mean molecular weight $\mu$ of the liquid above the crystal core (at a radius $r_{\rm cryst}$), which is then mixed by convection up to an outer radius $r_{\rm out}>r_{\rm cryst}$, where the ambient unperturbed $\mu$ is low enough to stabilize the fluid. The phase separation and mixing are calculated using the \textsc{mesa} stellar evolution code  \citep{Paxton2011,Paxton2013,Paxton2015,Paxton2018,Paxton2019,Jermyn2023}, with its recently implemented phase separation scheme \citep{Bauer2023}; see Section \ref{sec:scheme} for details.

Fig. \ref{fig:theory} shows how $r_{\rm out}$ expands as crystallization progresses, and how this expansion depends on the white dwarf's internal structure. Convective mixing during core helium burning in the white dwarf's progenitor star leaves behind an inner homogeneous region with uniform carbon and oxygen abundances, beyond which there is a small but sharp step in $\mu$ \citep{Salaris97,Straniero2003}. At early stages of crystallization, when $m_{\rm cryst}$ is small and the total mass of oxygen (carbon) that the crystal core has absorbed from (released to) the surrounding liquid is small, the liquid's $\mu$ is only slightly reduced. In this case, convection stops at the small $\mu$ step that corresponds to the jump in the C/O ratio (at $r_{\rm out}/R\approx 0.5$). Later on, as $m_{\rm cryst}$ grows, $\mu$ above the core decreases further such that this step can no longer stabilize convection and $r_{\rm out}$ expands until it reaches the white dwarf's hydrogen and helium atmosphere, where the ambient $\mu$ drops sharply again.  

\citet{Isern2017} suggested that the convective region between $r_{\rm cryst}$ and $r_{\rm out}$ can sustain a magnetic dynamo, analogous to the Earth's geodynamo. However, the convective velocity, and how the magnetic field scales with that velocity and with the white dwarf's rotation, are both uncertain \citep[e.g.][]{Christensen2010,Augustson2019,Schreiber2021Nat,Brun2022,Ginzburg2022,Fuentes2023,MontgomeryDunlap2023}. Here, we focus on an orthogonal question -- when can such magnetic fields reach the white dwarf's surface, where they may be observed? In Fig. \ref{fig:theory} we plot the magnetic diffusion time from $r_{\rm out}$ to the white dwarf's surface $R$, which is given by 
\begin{equation}\label{eq:tdiff}
    t_{\rm diff}=\int_{r_{\rm out}}^{R}{\frac{{\rm d}(r-r_{\rm out})^2}{\eta(r)}=\int_{r_{\rm out}}^{R}\frac{2(r-r_{\rm out}){\rm d}r}{\eta(r)}}.
\end{equation}
The magnetic diffusivity $\eta(r)=c^2/(4\upi\sigma)$ at each radius $r$ is computed similarly to \citet{Cantiello2016}, by interpolating between expressions that are valid in the non-degenerate, partially degenerate, and fully degenerate regimes \citep{Spitzer1962,Nandkumar1984,Wendell1987}; $c$ is the speed of light and $\sigma(r)$ is the electric conductivity.

As seen in Fig. \ref{fig:theory}, as long as $r_{\rm out}$ is stuck at the small $\mu$ step that is deep inside the CO core, the diffusion time is prohibitively long $t_{\rm diff}\sim 10\textrm{ Gyr}$. Only once $r_{\rm out}$ expands towards the white dwarf's hydrogen and helium atmosphere (when $m_{\rm cryst}$ grows sufficiently), $t_{\rm diff}$ shortens and enables the magnetic field to reach the surface. Specifically, the magnetic field's breakout time (age) $t=t_{\rm break}$ is found by solving
\begin{equation}\label{eq:breakout_condition}
 t_{\rm diff}(t)=t-t_{\rm cryst},
\end{equation}
where $t_{\rm cryst}$ marks the onset of crystallization in the white dwarf's centre (i.e. breakout is defined by when diffusion becomes faster than crystallization; see Section \ref{sec:scheme} for details).

We note that other processes, such as advection \citep{CharbonneauMacGregor2001} or magnetic buoyancy \citep{MacGregorCassinelli2003,MacDonaldMullan2004}, may in principle transport magnetic fields from $r_{\rm out}$ to the surface faster than diffusion. These mechanisms have been invoked -- with limited success -- for massive stars, where magnetic fields generated in the convective core have to penetrate through the radiative envelope to be observed. Their efficiency in the white dwarf context has not been studied yet, and we assume here that diffusion is the sole transport mechanism.  

\section{Computational scheme}\label{sec:scheme}

We construct a set of CO white dwarfs with masses $M\approx 0.5-1.0\,{\rm M}_{\sun}$ similarly to \citet{Bauer2023}, by using the \textsc{mesa} test suite \texttt{make\_co\_wd}, which evolves the progenitor $\approx 2.6-6.7\,{\rm M}_{\sun}$ stars through the various stages of stellar evolution, starting from the pre-main sequence. 
We then calculate the cooling and crystallization of these white dwarfs using the \texttt{wd\_cool\_0.6M} test suite. 
We use the \textsc{mesa} version r23.05.1, which implements the `Skye' equation of state \citep{Jermyn2021} and the new CO phase separation scheme of \citet{Bauer2023}. The carbon and oxygen profiles before and after phase separation can be seen in fig. 8 of \cite{Bauer2023}. 

Skye determines the phase (crystal or liquid) by minimizing the Helmholtz free energy, such that the critical $\Gamma_{\rm crit}$ for crystallization varies between $\Gamma_{\rm crit}\approx 200-230$, as a function of the C/O ratio \citep{Jermyn2021}. The crystallization time $t_{\rm cryst}$ is defined by the phase transition of the innermost cell. In the following time steps, the \texttt{phase\_separation} scheme propagates the crystallization front $r_{\rm cryst}$ outwards as the white dwarf cools and more cells crystallize. For each crystallizing cell, the scheme increases the oxygen mass fraction $X_{\rm O}$ and reduces the carbon mass fraction $X_{\rm C}\approx 1-X_{\rm O}$ by $\Delta X$, which is a function of $X_{\rm O}$, given by the \citet{BlouinDaligault2021} phase diagram (which has a similar $\Gamma_{\rm crit}$ to Skye). The total amounts of carbon and oxygen in the star are conserved by adjusting the composition of the cells above $r_{\rm cryst}$. This liquid becomes carbon-enriched and oxygen-depleted (i.e. opposite to the crystal core), triggering convection. At each time step, the scheme propagates the outer radius of the convection zone $r_{\rm out}$ iteratively, by mixing the cells at radii $r_{\rm cryst}<r<r_{\rm out}$ until the $\mu$ profile satisfies the Ledoux criterion for stability \citep[see][for details]{Bauer2023}.
We note that, in principle, thermohaline mixing may operate in regions that are stable according to the Ledoux criterion, but where $\nabla \mu>0$ \citep[e.g.][]{Paxton2013,Fuentes2023}. In practice, however, the Ledoux criterion for stability is reduced in our case to $\nabla\mu<0$ (the Ledoux term $\propto\nabla\mu$ dominates the criterion) -- the role of the entropy gradient in stabilizing the fluid is negligible.

In a previous paper, \citet{Ginzburg2022} used a simplified post-processing scheme to evaluate $r_{\rm out}$. As described above, here we employ more accurate variable $\Delta X(X_{\rm O})$ and $\Gamma_{\rm crit}(X_{\rm O})$ instead of their constant $\Delta X=0.2$ and $\Gamma_{\rm crit}=230$. More crucially, the white dwarfs in \citet{Ginzburg2022} do not experience phase separation as they crystallize. Instead, at each time $t$, $r_{\rm out}(t)$ is found by assuming that the \textit{unperturbed} liquid profile at $r>r_{\rm cryst}(t)$ is enriched by a carbon mass of $\Delta X m_{\rm cryst}(t)$, and depleted by a similar amount of oxygen. As demonstrated in our Fig. \ref{fig:theory} and in \citet{Bauer2023}, phase separation gradually changes the liquid carbon and oxygen profiles as crystallization progresses, leading to a difference between the two methods -- our current calculation with built-in phase separation is more precise.   

The magnetic diffusion time from $r_{\rm out}$ to the surface is calculated using equation \eqref{eq:tdiff}, and the breakout time $t=t_{\rm break}$ is found in accordance with equation \eqref{eq:breakout_condition}, when the condition $t_{\rm diff}(t)<t-t_{\rm cryst}$ is first met. These equations provide a good estimate for $t_{\rm break}$ at high white dwarf masses, for which $t_{\rm diff}(t)$ drops very sharply. At lower white dwarf masses, $t_{\rm diff}(t)$ declines more gradually (Fig. \ref{fig:theory}), necessitating the incorporation of a time-dependent magnetic diffusion solver in \textsc{mesa}, which is beyond the scope of this work. This difference with $M$ stems from the structure of the initial (i.e. before crystallization) C and O profiles beyond the original C/O jump (Fig. \ref{fig:theory}). As seen in fig. 8 of \citet{Bauer2023}, these profiles are flatter at high $M$, leading to a sharper transition of $r_{\rm out}(t)$ from the small $\mu$ step (that corresponds to the C/O jump) to the larger $\mu$ drop at the helium layer -- where the diffusion time is much shorter (Fig. \ref{fig:theory}).

\section{Results}\label{sec:results}

\begin{figure}
\includegraphics[width=\columnwidth]{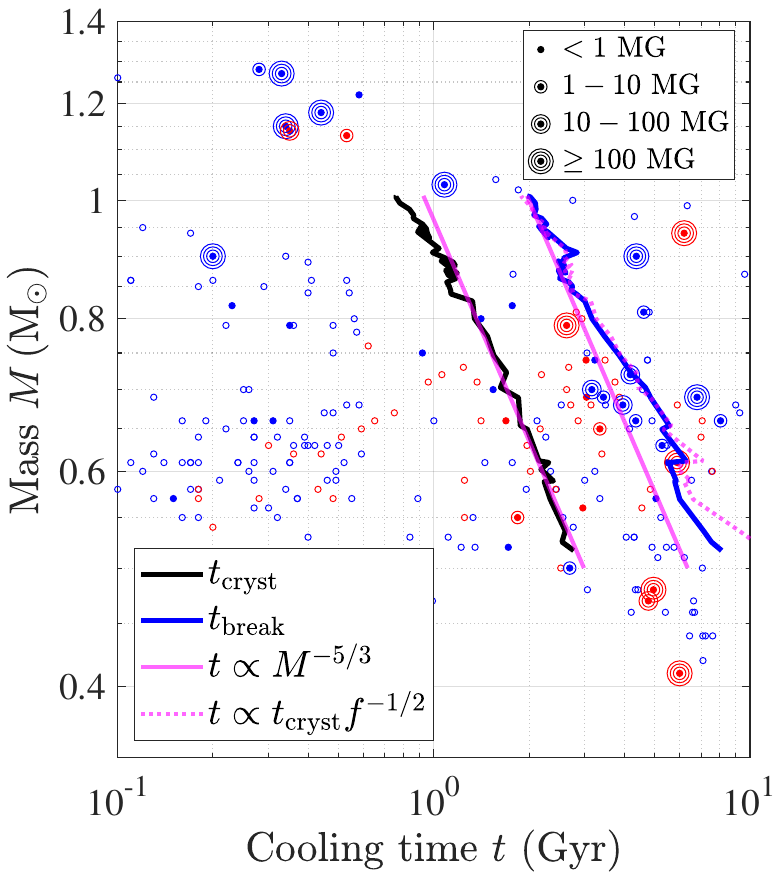}
\caption{The onset of crystallization $t_{\rm cryst}$, and the magnetic field's breakout to the surface $t_{\rm break}$, as a function of the white dwarf's mass $M$. $t_{\rm cryst}$ is approximated well by equation \eqref{eq:t_cryst}, whereas the breakout time is fit better by equation \eqref{eq:t_f}: $t_{\rm break}\propto t_{\rm cryst}f^{-1/2}$, with the mass fraction $f$ that remains uncrystallized at breakout extracted from \textsc{mesa} (plotted in Fig. \ref{fig:fraction}). The observed white dwarfs (blue for hydrogen atmospheres and red for helium) are from the volume-limited sample of \citet{Bagnulo_2022}. Empty dots represent non-magnetic white dwarfs and filled dots mark magnetic ones, with the strength of the measured magnetic field indicated by the number of surrounding circles.}
\label{fig:obs}
\end{figure}

\begin{figure}
\includegraphics[width=\columnwidth]{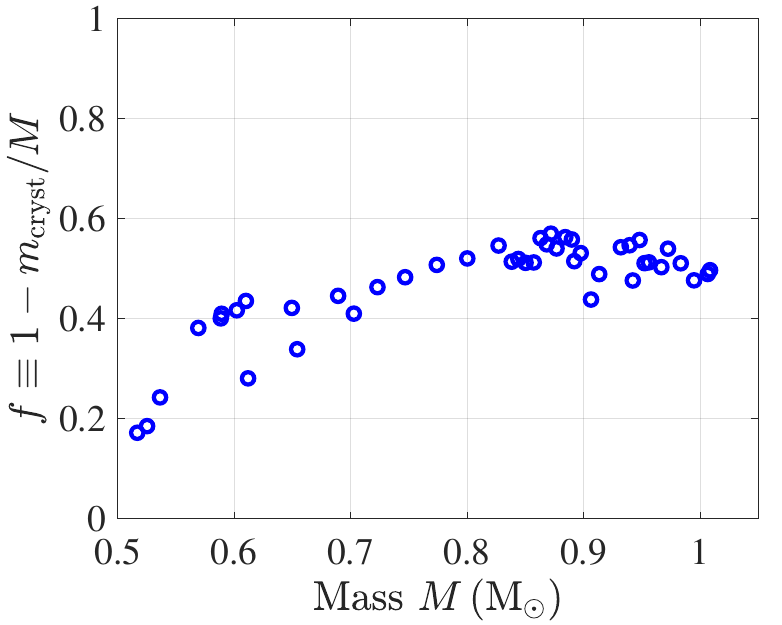}
\caption{
The white dwarf's uncrystallized mass fraction $f$ at the moment of the magnetic field's breakout to the surface $t_{\rm break}$. In lower mass white dwarfs, a larger fraction $m_{\rm cryst}/M=1-f$ has to crystallize before the field breaks out, leading to a deviation from a $t_{\rm break}\propto M^{-5/3}$ scaling (Fig. \ref{fig:obs}).
}
\label{fig:fraction}
\end{figure}

Fig. \ref{fig:obs} shows our computed crystallization and breakout times as a function of the white dwarf's mass $M$. The crystallization times follow the analytical model of \citet{Mestel1952} -- $t_{\rm cryst}\propto M^{-5/3}$. The breakout times $t_{\rm break}$ are significantly longer: roughly half of the white dwarf's mass has to crystallize before the magnetic field emerges to the surface (Fig. \ref{fig:fraction}), with an even higher fraction $m_{\rm cryst}/M\equiv 1-f$ at low masses $M$ \citep[because of the difference in the initial carbon and oxygen profiles; see fig. 8 in][]{Bauer2023}. 
This variation in $f(M)$ explains the deviation from a $t_{\rm break}\propto M^{-5/3}$ scaling, as implied by equation \eqref{eq:t_f}, but with an order-unity fitting coefficient (though formally our $f\ll 1$ approximation breaks down).

\citet{Bagnulo_2022} compared their volume-limited white dwarf sample to $t_{\rm cryst}(M)$. They concluded that massive white dwarfs ($M\gtrsim 1.0\,{\rm M}_{\sun}$) were likely magnetized by mergers, with strong magnetic fields often appearing before crystallization \citep[see also][]{Caiazzo2021,Fleury2022}. Such massive white dwarfs are beyond the scope of our CO crystallization dynamo model because they may harbour oxygen--neon cores \citep[e.g.][]{Camisassa2022}. Similarly, $M\lesssim 0.5\,{\rm M}_{\sun}$ white dwarfs potentially harbour helium cores and are thus also beyond the scope of this work \citep[e.g.][]{PradaMoroni2009}. In the relevant $\approx 0.5-1.0\,{\rm M}_{\sun}$ range, \citet{Bagnulo_2022} found that the frequency and strength of magnetic fields increase after the crystallization line is crossed, signifying the potential contribution of the crystallization dynamo mechanism \citep[see also][]{Amorim2023,Caron2023,Hardy2023}. 

However, the breakout time $t_{\rm break}(M)>t_{\rm cryst}(M)$ provides a more accurate test of the crystallization dynamo channel, because it accounts for the magnetic diffusion to the surface. As seen in Fig. \ref{fig:obs}, the difference between crystallization and breakout is of order a few Gyr, potentially explaining the observed delay between crystallization and the appearance of strong magnetic fields -- $t_{\rm break}$ fits better the age at which magnetic fields $B\gtrsim 1\textrm{ MG}$ first emerge.

\citet{Bagnulo_2022} made the distinction between white dwarfs with thick hydrogen envelopes (blue symbols in Fig. \ref{fig:obs}) and hydrogen-deficient white dwarfs, which exhibit helium spectra (red symbols). The difference in $t_{\rm cryst}$ between the two spectral types (due to their different cooling rates) can be seen in fig. 2 of \citet{Bagnulo_2022}.
All our simulated white dwarfs possess thick hydrogen envelopes \citep[$10^{-6}-10^{-4}\,{\rm M}_{\sun}$; see][for comparison]{Bedard2020}. We do not calculate $t_{\rm break}$ for white dwarfs with helium atmospheres, and focus instead on the sensitivity to other parameters that influence the breakout process more directly.

\subsection{Phase diagram}
\label{subsec:phase_diagram}

\begin{figure}
\includegraphics[width=\columnwidth]{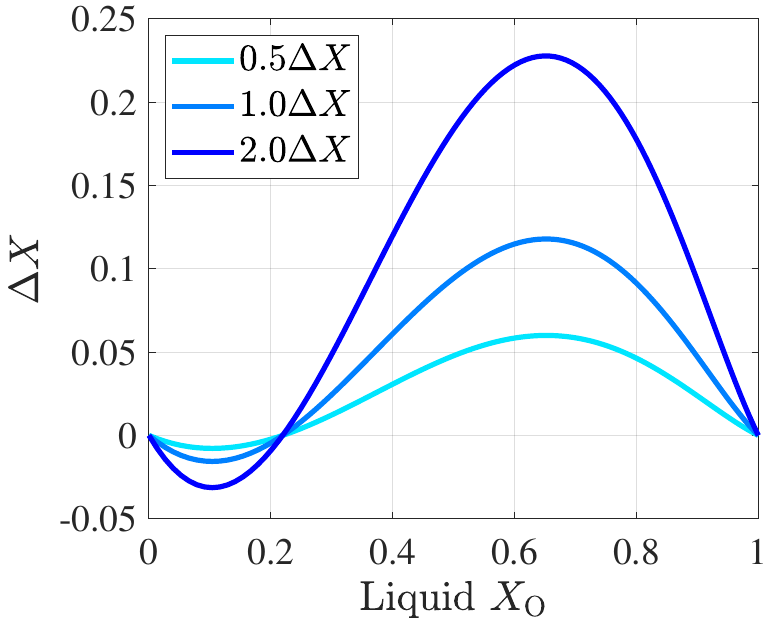}
\caption{Oxygen mass fraction enrichment $\Delta X$ at the phase transition from liquid to crystal, as a function of the liquid $X_{\rm O}$. The nominal $\Delta X$ is taken from \citet{BlouinDaligault2021}, after converting from number fractions $x$ to mass fractions $X$. Changing $\Delta x$ by a factor of 2 in each direction is almost identical to changing $\Delta X$, and it roughly bounds the variation between different phase diagrams \citep[see fig. 6 in][]{BlouinDaligault2021}.}
\label{fig:blouin}
\end{figure}

The degree of oxygen enrichment $\Delta X$ during crystallization is a key factor in determining the delay between crystallization and breakout. A higher $\Delta X$ requires more oxygen to be drawn from the surrounding liquid to form a crystal core with a given $m_{\rm cryst}$, reducing the liquid's $\mu$ faster. As can be understood from Fig. \ref{fig:theory}, this faster reduction in $\mu$ leads to a more rapid expansion of $r_{\rm out}$, and thereby to an earlier breakout of the magnetic field. Our nominal $\Delta X(X_{\rm O})$, given by the phase diagram of \cite{BlouinDaligault2021}, is plotted in Fig. \ref{fig:blouin}.

\begin{figure}
\includegraphics[width=\columnwidth]{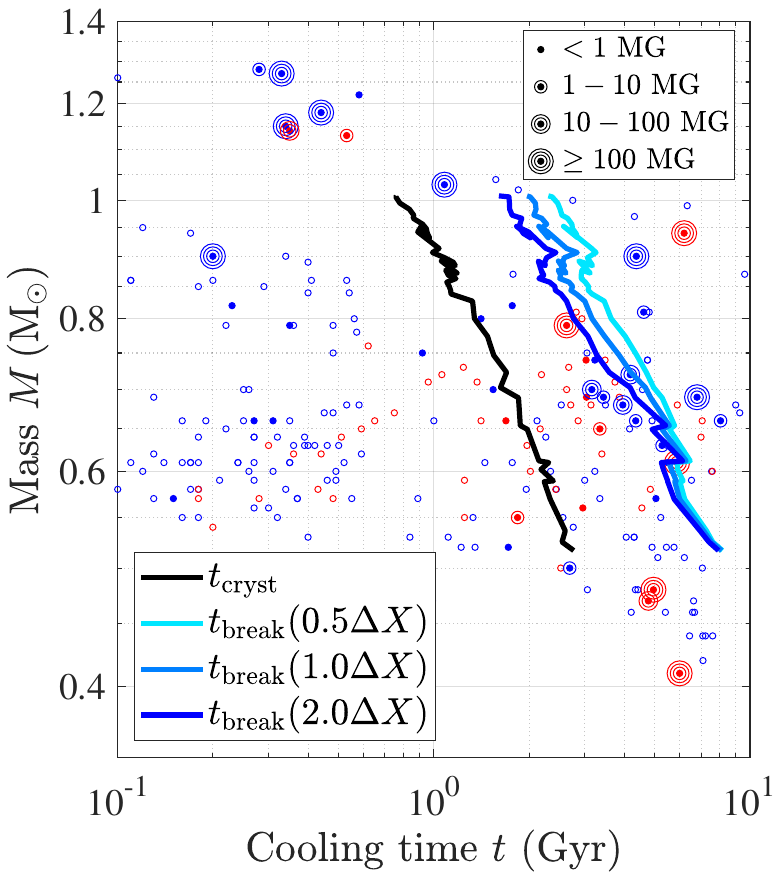}
\caption{The sensitivity of $t_{\rm break}$ to the uncertainty in the CO phase diagram. The breakout process depends on the degree of oxygen enrichment $\Delta X$ during crystallization, which is plotted in Fig. \ref{fig:blouin}. The onset of crystallization $t_{\rm cryst}$ is independent of $\Delta X$, and we do not vary the crystallization curve $\Gamma_{\rm crit}(X_{\rm O})$.}
\label{fig:phase}
\end{figure}

In Fig. \ref{fig:phase} we test the dependence of $t_{\rm break}$ on the uncertainty in the phase diagram by varying $\Delta X$ by a factor of 2 in each direction (more precisely, we change the enrichment in the number fraction $\Delta x$, which is almost identical to changing $\Delta X$; see Fig. \ref{fig:blouin}). This variation roughly represents the difference between \citet{BlouinDaligault2021} and other phase diagrams \citep{Horowitz2010,MedinCumming2010}. In our $M\approx 0.5-1.0\,{\rm M}_{\sun}$ white dwarfs, the initial central liquid $X_{\rm O}\approx 0.6-0.7$ \citep{Bauer2023}, for which the nominal $\Delta X\approx 0.1$. The interplay between this $\Delta X$ and the $X_{\rm O}$ step (of comparable height) at the edge of the inner homogeneous region in the initial profile sets $t_{\rm break}$ (Fig. \ref{fig:theory}). The height and location of this step change with $M$ \citep[fig. 8 of][]{Bauer2023}, possibly explaining the converging $t_{\rm break}$ at low $M$ in Fig. \ref{fig:phase}.

In principle, an additional uncertainty in $\Gamma_{\rm crit}(X_{\rm O})$ of the phase diagram would shift both $t_{\rm cryst}$ and $t_{\rm break}$; see e.g. fig. 7 in \citet{Jermyn2021} for the (small) difference between Skye (our $\Gamma_{\rm crit}$) and \citet{BlouinDaligault2021}, which is the same as \citet{Blouin2020}. In this paper, however, we choose to focus on $\Delta X$ because it directly changes the delay time $\Delta t\equiv t_{\rm break}-t_{\rm cryst}$. 

\subsection{Reaction rates} \label{subsec:reaction_rate}

The second key factor in determining $t_{\rm break}$, as can be appreciated from Fig. \ref{fig:theory}, is the initial C/O profile before crystallization. Specifically, the breakout time is sensitive to the details of the white dwarf's inner homogeneous zone: its radial extent, the central C/O ratio, and the C/O jump beyond this region (followed by a more gradual rise in the C/O ratio). All these features are set during the core helium burning phase of the white dwarf's progenitor star \citep{Salaris97,Salaris2010,Straniero2003}. The magnetic field's breakout to the surface may therefore probe the various uncertainties of this stellar evolution phase, such as convective overshooting and the $3\alpha$ and $^{12}{\rm C}(\alpha,\gamma)^{16}{\rm O}$ nuclear reactions, which synthesize carbon and oxygen \citep[e.g.][]{Chidester2023}.

As a specific example, we focus here on the poorly constrained $\alpha$ capture reaction $^{12}{\rm C}(\alpha,\gamma)^{16}{\rm O}$, which converts carbon into oxygen \citep{DeBoer2017}. This reaction is of great astrophysical importance, and it has recently gained attention in the context of LIGO's detection of gravitational waves from binary black hole mergers. Pair-instability supernova theory predicts a gap in the black hole mass distribution at $\sim 50-130\,{\rm M}_{\sun}$ \citep[e.g.][]{Belczynski2016,Woosley2017,WoosleyHeger2021}. However, the location of this gap is sensitive to the $^{12}{\rm C}(\alpha,\gamma)^{16}{\rm O}$ reaction rate \citep{Farmer2019,Farmer2020,Shen2023} -- leading to uncertainties in interpreting the results from LIGO/Virgo/KAGRA.  

\begin{figure}
\includegraphics[width=\columnwidth]{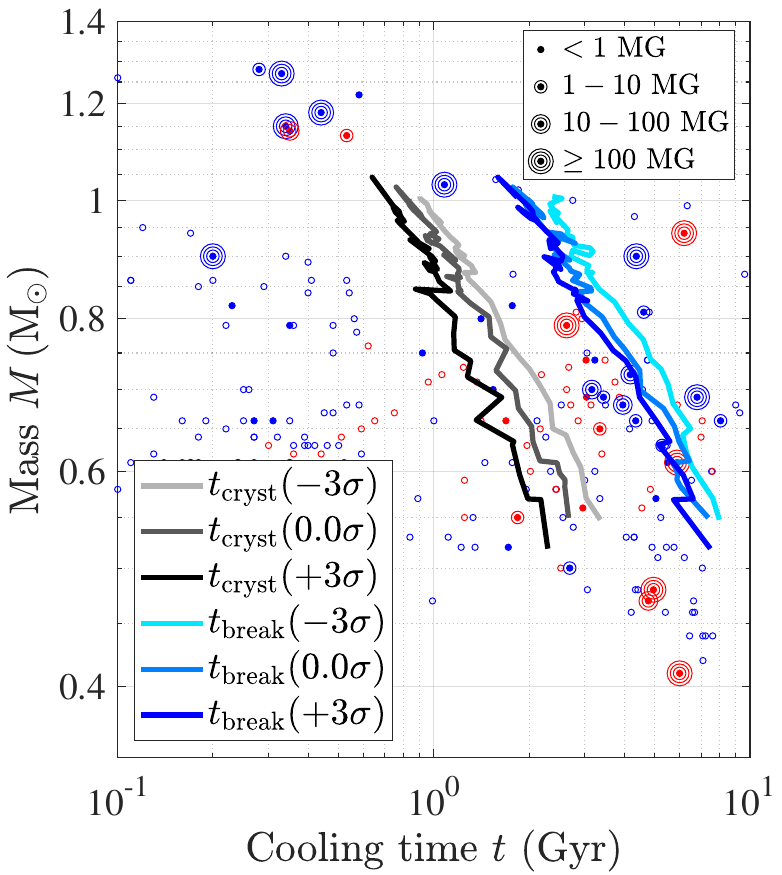}
\caption{The sensitivity of $t_{\rm cryst}$ and $t_{\rm break}$ to the uncertainty in the $^{12}{\rm C}(\alpha,\gamma)^{16}{\rm O}$ nuclear reaction rate during core helium burning in the white dwarf's progenitor star. Nominal and $\pm 3\sigma$ values are from fig. 9 of \citet{Mehta2022}, who expanded the \citet{DeBoer2017} rates \citep[see also fig. 1 in][]{Chidester2022}. The nominal rates produce similar curves to Fig. \ref{fig:obs}, where we used the older \citet{Kunz2002} reaction rates.}
\label{fig:change_sigma}
\end{figure}

In Fig. \ref{fig:change_sigma} we show how our results depend on the $^{12}{\rm C}(\alpha,\gamma)^{16}{\rm O}$ reaction rate during core helium burning. We followed \citet{Chidester2022} and used the \citet{DeBoer2017} reaction rates that were expanded by \citet{Mehta2022} to include $\pm 3\sigma$ uncertainties. The nominal \citet{DeBoer2017} rates produce $t_{\rm cryst}$ and $t_{\rm break}$ that are very similar to the ones obtained in previous sections, using the older \citet{Kunz2002} rates \citep[for consistency with][]{Bauer2023}. When using the $\pm 3 \sigma$ rates, however, \texttt{make\_co\_wd} produces white dwarfs with different carbon and oxygen distributions -- changing both the crystallization and breakout times. 

\subsection{Metallicity}\label{sec:metal}

\begin{figure}
\includegraphics[width=\columnwidth]{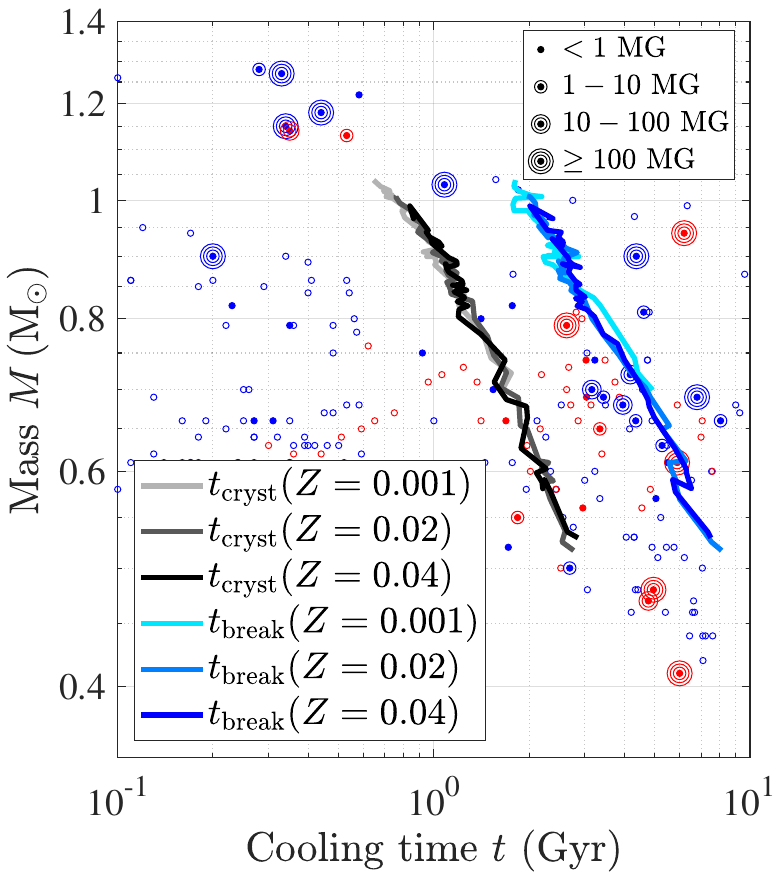}
\caption{The sensitivity of $t_{\rm cryst}$ and $t_{\rm break}$ to the progenitor star's initial metallicity $Z$. In previous plots we used the nominal $Z=0.02$. Sedimentation and possible distillation of $^{22}{\rm Ne}$, which are sensitive to $Z$, are not considered in the plot and are discussed in the text. For this plot, we used the \citet{Kunz2002} nuclear reaction rates.}
\label{fig:metal}
\end{figure}

The carbon and oxygen profiles may also depend on the progenitor star's initial metallicity $Z$. However, as seen in Fig. \ref{fig:metal}, while the white dwarf's final mass $M$ is sensitive to its progenitor's metallicity, the $t_{\rm cryst}(M)$ and $t_{\rm break}(M)$ curves are largely unchanged for metal-poor ($Z=0.001)$ and metal-rich ($Z=0.04)$ stars.

We emphasize that our models do not include the effect of gravitational energy release by $^{22}{\rm Ne}$ sedimentation on the white dwarf's cooling \citep{BildstenHall2001}. For $M=0.6\,{\rm M}_{\sun}$ and our nominal $Z=0.02$, \citet{Bauer2023} finds a modest cooling delay due to sedimentation of $\approx 0.3$ Gyr, which is below the noise level of our $t_{\rm break}(M)$ curves. However, recently observed cooling delays for some white dwarfs suggest that $^{22}{\rm Ne}$ settling might be more significant  
\citep{Althaus2010,Cheng2019,Bauer2020,Camisassa2021}. Specifically, in Section \ref{sec:distillation} we discuss the distillation process, which may transport $^{22}{\rm Ne}$ towards the centre more efficiently \citep[e.g.][]{Blouin2021}. The dependence of such mechanisms on the $^{22}{\rm Ne}$ abundance implies a much larger -- yet currently uncertain -- sensitivity to $Z$ than Fig. \ref{fig:metal} suggests. We defer the investigation of the influence of $^{22}{\rm Ne}$ to future work. 

\section{Summary and discussion}\label{sec:summary}

Convective dynamos operating during crystallization present a promising channel to white dwarf magnetism \citep{Isern2017}. Even though the magnitude of the generated magnetic fields is uncertain \citep{Schreiber2021Nat,Ginzburg2022,Fuentes2023,MontgomeryDunlap2023}, their timing is an orthogonal question that may be used to constrain the relative contribution of this channel. The dynamo operates in a convective mantle that surrounds the crystallizing core. In order to be observed, the generated magnetic field has to diffuse out from this mantle to the white dwarf's surface through convectively-stable layers. The delay time -- due to diffusion -- between the onset of crystallization $t_{\rm cryst}$ and the breakout of the magnetic field to the surface $t_{\rm break}$ is the focus of our paper.

We computed the crystallization and breakout times in CO white dwarfs with masses $M\approx 0.5 - 1.0\,{\rm M}_{\sun}$, using the phase separation scheme that has been recently implemented in \textsc{mesa} \citep{Bauer2023}. This scheme utilizes the Skye equation of state \citep{Jermyn2021} and the \citet{BlouinDaligault2021} phase diagram to separate the CO mixture into an oxygen-rich crystal and a carbon-rich liquid which is unstable to convection due to its reduced molecular weight $\mu$. The scheme updates the $\mu(r)$ profile iteratively to determine the outer radius of the convection $r_{\rm out}$, from which we calculate the magnetic diffusion time to the surface $t_{\rm diff}$. 

As crystallization progresses, and a larger fraction $m_{\rm cryst}/M$ of the white dwarf crystallizes, $r_{\rm out}$ expands in two distinct stages (Fig. \ref{fig:theory}). At early times, convection is stabilized by a small $\mu$ step inside the CO core -- where the C/O ratio jumps -- a relic of core helium burning and convective mixing in the white dwarf's progenitor star. At later times, as $m_{\rm cryst}$ grows and $\mu$ of the surrounding liquid is reduced further, convection extends all the way to the white dwarf's helium layer. The magnetic diffusion time changes dramatically between these two stages. As long as $r_{\rm out}$ is deep inside the CO core, $t_{\rm diff}\gtrsim 10\textrm{ Gyr}$ -- trapping the magnetic field. Once $r_{\rm out}$ reaches the helium layer, on the other hand, $t_{\rm diff}\lesssim 1\textrm{ Gyr}$ -- enabling the field to reach the surface. At high masses $M$, the transition between these regimes is sharp, with a well defined breakout time $t_{\rm break}$. At lower $M$, the transition is more gradual, and we approximately estimate $t_{\rm break}\approx t_{\rm cryst}+t_{\rm diff}(t_{\rm break})$, i.e. when the diffusion time becomes shorter than the time elapsed since the onset of crystallization.   

Our computed $t_{\rm cryst}$ and $t_{\rm break}$ are presented in Fig. \ref{fig:obs}, where they are compared to the \citet{Bagnulo_2022} volume-limited white dwarf sample. The crystallization times are fit well by $t_{\rm cryst}\propto M^{-5/3}$, which is derived from the analytical cooling theory of \citet{Mestel1952}. The breakout times deviate from this scaling and are fit better by the generalized $t_{\rm break}\propto t_{\rm cryst}f^{-1/2}$, where $f\equiv1-m_{\rm cryst}/M$. In other words, a somewhat different fraction $m_{\rm cryst}/M$ has to crystallize in white dwarfs with different masses $M$ before the magnetic field breaks out to the surface (a result of the difference in initial C/O profiles before crystallization). 

As seen in Fig. \ref{fig:obs}, strong magnetic fields $B\gtrsim 1\textrm{ MG}$ first appear in proximity to $t_{\rm break}(M)$ for $M\approx 0.5 - 1.0\,{\rm M}_{\sun}$, except for a single $0.9\,{\rm M}_{\sun}$ outlier -- ostensibly signifying the dominant contribution of the crystallization dynamo channel to strongly magnetic CO white dwarfs. Lower-mass white dwarfs may contain helium cores, whereas higher-mass white dwarfs may contain ONe cores, or might have been magnetized by mergers -- both of these white dwarf classes are beyond the scope of this work.
Given the potential challenges of the crystallization dynamo theory \citep{Fuentes2023,MontgomeryDunlap2023}, it is noteworthy that even if crystallization cannot sustain a strong magnetic dynamo on its own, the triggered convection may help to convey existing (``fossil'') fields to the white dwarf's surface. In order to explain the observed late appearance of strong magnetism, such fossil fields must initially be buried deep inside the CO core, because the magnetic diffusion time from the edge of the CO core to the surface is shorter than a Gyr (Fig. \ref{fig:theory}). If the fossil field is confined to deeper layers of the CO core (where the diffusion time is much longer), then it must wait for crystallization-driven convection to transport it closer to the surface -- where it will emerge at a similar time to our computed $t_{\rm break}$. For example, a convective dynamo operating during core helium burning might leave behind a fossil magnetic field that is confined to about half of the CO core's radius (the inner initially homogeneous region of the profiles in Fig. \ref{fig:theory}, which is a relic of convection during this stage).

Nevertheless, several of the observed magnetic white dwarfs with $M\approx 0.7\,{\rm M}_{\sun}$ are in fact slightly younger than $t_{\rm break}$, challenging the crystallization dynamo (or a related) channel. This result seems to hold even when considering the uncertainties in the phase diagram (Fig. \ref{fig:phase}) and in the nuclear reaction rates (Fig. \ref{fig:change_sigma}).  
Besides the possibility that these white dwarfs were actually magnetized by a different process, we identify two potential reasons for their offset from $t_{\rm break}(M)$. One source of uncertainty is convective overshooting -- penetration and mixing of material beyond a convection zone \citep[see][for recent progress]{Anders2022,Jermyn2022,Blouin2023,Blouin2024}. \citet{Chidester2023} demonstrated that overshooting during core helium burning in the white dwarf's progenitor star may significantly alter the initial C/O profiles before crystallization, which may in turn shift $t_{\rm break}$ once crystallization unfolds. Calculating the effects of overshooting is beyond the scope of our work, because the non-monotonic initial C and O profiles in that case \citep[fig. 1 of][]{Chidester2023} complicate our analysis. Another source of inaccuracy is our definition of $t_{\rm break}$. For $M\approx 0.8\,{\rm M}_{\sun}$, $t_{\rm diff}(t)$ drops quickly, such that the breakout time is well defined. For $M\approx 0.6\,{\rm M}_{\sun}$, $t_{\rm diff}(t)$ declines more gradually (Fig. \ref{fig:theory}), leading to inconsistencies in our analysis. A consistent solution requires the incorporation of a magnetic diffusion solver in \textsc{mesa} -- instead of equation \eqref{eq:tdiff} -- which is also beyond the scope of the current paper.

While our technique may be used to constrain the relative role of different magnetization channels, we can also turn the argument on its head and use white dwarf magnetism as a tool. If we assume that crystallization dynamos are indeed the main channel to produce $\gtrsim 1\textrm{ MG}$ magnetic fields in $M\approx 0.5-1\,{\rm M}_{\sun}$ white dwarfs, then the observed $t_{\rm break}(M)$ -- i.e. when strong fields first appear -- may be used to study several physical processes that govern the formation and evolution of CO white dwarfs. Specifically, the breakout time $t_{\rm break}$ and the mass fraction that crystallizes at this time $m_{\rm cryst}/M\equiv 1-f$ are sensitive to the initial C/O profile before crystallization, as well as to the CO phase diagram which determines how this profile changes during crystallization. 

The dependence of our results on the phase diagram is estimated in Fig. \ref{fig:phase}. In Fig. \ref{fig:change_sigma} we show the dependence on the poorly constrained $^{12}{\rm C}(\alpha,\gamma)^{16}{\rm O}$ nuclear reaction rate, which together with the $3\alpha$ reaction and convective overshooting sets the white dwarf's C/O profile during core helium burning \citep[e.g.][]{Salaris2010,Chidester2022,Chidester2023,Tognini2023}. While the variations that we find in $t_{\rm break}$ are relatively small, a large enough white dwarf survey may in principle utilize $t_{\rm break}(M)$ to distinguish between different theoretical models. Thus, white dwarf magnetism may complement asteroseismology as an independent tool to probe the internal structure of white dwarfs \citep{Metcalfe2001,Metcalfe2002,Metcalfe2003}. Interestingly, the seismic signature is sensitive to similar features of the C/O profile as the breakout process, namely the region between the jump in the C/O ratio and the helium layer 
\citep{Chidester2022,Chidester2023}. 
Specifically, several asteroseismic studies indicate that the oxygen-rich central homogeneous zone is larger than predicted by standard stellar evolution models \citep{Giammichele2018,Giammichele2022}, possibly due to extra convective boundary mixing during the core helium burning phase \citep[see also][]{Constantino2015,Paxton2018}. The timing of white dwarf magnetism could provide additional constraints on such mixing mechanisms.

As demonstrated by \citet{Farmer2020}, the $^{12}{\rm C}(\alpha,\gamma)^{16}{\rm O}$ reaction rate also determines the location of the pair-instability supernova mass gap, which is probed by LIGO/Virgo/KAGRA observations of merging black holes \citep{LIGO2019,LIGO2023}. A more accurate determination of $t_{\rm break}$ (again, assuming that the crystallization dynamo is the dominant magnetization channel of CO white dwarfs) might therefore shed some light on these gravitational wave observations.

\subsection{Distillation of minor species}\label{sec:distillation}

This paper sets the stage for studying the related process of distillation. Despite their low abundance, neutron-rich trace elements such as $^{22}{\rm Ne}$ and $^{56}{\rm Fe}$ may significantly alter the crystallization process \citep{Isern1991,Segretain1994,Segretain1996,Blouin2021,Caplan2023}. If the solid crystals are sufficiently depleted in such elements compared to the liquid, then they float upwards -- away from the crystallization front -- until they melt and mix in the lower density environment. This process gradually enriches the liquid at the bottom, until it crystallizes later at a higher $\Gamma_{\rm crit}$ (due to the different composition) -- forming a crystal layer enriched in these trace elements (i.e. crystallization ``distills'' the trace elements). Recently, \citet{Blouin2021} demonstrated that the gravitational energy release by $^{22}{\rm Ne}$ distillation may explain the cooling delays observed for a fraction of massive white dwarfs \citep[see also][]{KenShen2023}. 

In principle, the convective motions during the distillation process may sustain a magnetic dynamo. The large amount of released gravitational energy suggests that such distillation-driven dynamos might even generate stronger magnetic fields than classical crystallization-driven dynamos. In order to compute the breakout time of these magnetic fields to the surface, $r_{\rm out}$ of the convection has to be recalculated using a three-component phase diagram and phase separation scheme that include $^{22}{\rm Ne}$. Such a calculation is beyond the scope of this paper.

According to \citet{Blouin2021}, the distillation process is sensitive to the metallicity, and specifically to the $^{22}{\rm Ne}$ abundance. Interestingly, for a standard $^{22}{\rm Ne}$ abundance and for $X_{\rm O}=0.6$ (similar to our models), \citet{Blouin2021} find that distillation does not start at the onset of crystallization, but rather only when $m_{\rm cryst}/M\approx 0.6$ of the core has already crystallized. Coincidentally, this is similar to the mass fraction that has to crystallize before the magnetic field from a crystallization dynamo breaks out to the surface (Fig. \ref{fig:fraction}). We therefore conclude that -- if generated -- magnetic fields from distillation dynamos first appear close to our computed $t_{\rm break}$, potentially providing an alternative explanation for the observations. This preliminary conclusion motivates future work on distillation dynamos.

\section*{Acknowledgements}

We thank the \textsc{mesa} team for constantly improving this powerful open-source tool. Specifically, we thank Evan Bauer for implementing the phase separation scheme, which we found especially useful.
We also thank Evan Bauer, Simon Blouin, Jay Farihi, Jim Fuller, Na'ama Hallakoun, Selma de Mink, Ken Shen, and the anonymous reviewer for helpful discussions and comments which improved the paper. We acknowledge support from the Israel Ministry of Innovation, Science, and Technology (grant No. 1001572596), and from the United States -- Israel Binational Science Foundation (BSF; grant No. 2022175). \texttt{MesaScript} \citep{bill_wolf_mesascript} was used to automate some of the steps in this work.

%%%%%%%%%%%%%%%%%%%%%%%%%%%%%%%%%%%%%%%%%%%%%%%%%%
\section*{Data Availability}

The \textsc{mesa} input files required to reproduce our results are available at \url{https://zenodo.org/doi/10.5281/zenodo.10520429}.



\bibliographystyle{mnras}
%\bibliography{breakout} % if your bibtex file is called example.bib
\input{breakout.bbl}

%%%%%%%%%%%%%%%%%%%%%%%%%%%%%%%%%%%%%%%%%%%%%%%%%%

%%%%%%%%%%%%%%%%% APPENDICES %%%%%%%%%%%%%%%%%%%%%

%%%%%%%%%%%%%%%%%%%%%%%%%%%%%%%%%%%%%%%%%%%%%%%%%%

% Don't change these lines
\bsp	% typesetting comment
\label{lastpage}
\end{document}